\documentclass[review,1p,2times]{elsarticle}

\usepackage{tipa}
\usepackage[fleqn]{amsmath}
\usepackage{mathrsfs}
\usepackage{amssymb}
\usepackage{multirow}
\usepackage{graphicx}
\usepackage{subfigure}
\usepackage{booktabs}
\usepackage{amsthm,amsmath}
\usepackage{mathptmx}

\begin{document}

\begin{frontmatter}



\title{Estimating Linear Mixed-effects State
Space Model Based on Disturbance Smoothing}
 \tnotetext[]{Corresponding author, email: \textit{zhoujie@xidian.edu.cn.} Financial supports from
 NSFC (71271165) is gratefully acknowledged.}
\author{Jie Zhou $^{\star}$}
\author{Aiping Tang}

\address{Department of Statistics, Xidian University,
 Xi'an, 710071, P R China}
\begin{abstract}
We extend the linear mixed-effects state space model to accommodate
the correlated individuals and investigate its parameter and state
estimation based on disturbance smoothing in this paper. For
parameter estimation, EM and score based algorithms are considered.
Intermediate quantity of EM algorithm is investigated firstly from
which  the explicit recursive formulas for the maximizer of the
intermediate quantity are derived out for two given models. As for
score based  algorithms, explicit formulas for the score vector are
achieved from which it is shown that the maximum likelihood
estimation  is equivalent to   moment estimation. For state
estimation
 we advocate it should be carried out without assuming the random effects being
known in advance especially  when the  longitudinal observations are
sparse. To this end an algorithm named kernel smoothing based
mixture Kalman filter (MKF-KS) is proposed. Numerical studies are
carried out to investigate the proposed algorithms which validate
the efficacy of the proposed inference approaches.
\end{abstract}


\end{frontmatter}


\section{Introduction}\label{introduction}
State space models  are widely used in various fields such as
economics, engineering, biology et al. In  particular structural
time series models are just the special  state space models.  For
linear state space model with Gaussian error, it is known that
Kalman filter  is optimal for state estimation. For nonlinear state
space model, there does not exist optimal algorithm and various
suboptimal algorithms for state estimation  have been proposed in
literatures, see Harvey (1989), Durbin and Koopman (2012) for
details about these algorithms. Traditionally the state space models
are designed for the single processes.
\par In recent years in order to deal with the
longitudinal data, the state space models for the multiple processes
have been proposed and much attention has been attracted in this
field.  These models can be classified into two categories, i.e.,
the discrete  and continuous models. For the single processes the
discrete models are often  referred as the hidden Markov models
(HMMs).  Historically the discrete models with random effects were
introduced by Langeheine and van de Pol (1994) while Altman (2007)
provided a general framework for implementing the random effects in
the discrete models. For the parameter estimation, Altman (2007)
evaluated the likelihood as a product of matrixes and performed
numerical integration via Gaussian quadrature. A quasi-Newton method
is used for maximum likelihood estimation. Maruotti (2011) discussed
mixed hidden Markov models and their estimation using EM algorithm.
Jackson et al (2014) extended the work of Altman (2007) by allowing
the hidden state to jointly model longitudinal binary and count
data. The likelihood was evaluated by forward-backward algorithm and
adaptive Gaussian quadrature.  For continuous state space models,
Gamerman and Migon (1993) was the first to  use the state space
model to deal with multiple processes. They proposed  dynamic
hierarchical models for the longitudinal data. Unlike the usual
hierarchical model where the parameters are modeled by hierarchical
structure,   the hierarchy in Gamerman and Migon (1993) is built for
the state variables.  Landim and Gamerman (2000) generalized such
models to multiple processes. It should be noted that dynamic
hierarchical models are still the linear state space models with
Gaussian error and so the statistical inference for such model can
be carried out using the traditional method. Lodewyckx et al (2011)
proposed hierarchical linear state space model to model the emotion
dynamics. Here the hierarchy is built for the parameters. Unlike the
models in Gamerman and Migon (1993), these models are essentially
the nonlinear state space model and Baysian approach was employed to
estimate the unknown parameters. Liu et al (2011) proposed a similar
model, which was called mixed-effects state space model (MESSM), to
model the longitudinal observations of  a group of HIV infected
patients. As for the statistical inference of the model, both  EM
algorithm and Baysian approach were investigated. In order to
justify their statistical inference, Liu et al (2011) assumed that
the individuals in the group are independent and  the model should
have a linear form of parameter. As for the state estimation, they
took the predicted values of random effects as the true values and
then estimate the state using Kalman filter.
\par In this paper we extend the models proposed in Liu et al
(2011) and Lodewyckx et al (2011). The proposed models  can
accommodate the group with correlated individuals and do not require
 the models should possess  the linear form of parameters. The model will
 still
 be named as  MESSM just as in Liu et al (2011).
   For this generalized MESSM,  both the parameter and state estimation are
considered. As for  parameter estimation,  EM algorithm is firstly
considered. Unlike Liu et al (2011) in which EM algorithm is based
on state smoothing, we establish  the EM algorithm based  on the
disturbance smoothing which  greatly simplifies EM algorithm.
Actually the proposed EM algorithm can be seen  as the
Rao-Blackwellized version of that proposed in Liu et al (2011).  For
two important special MESSM's, we get the elegant recursive formula
for the maximizer  of intermediate quantity of EM algorithm.  Since
the convergence rate of EM algorithm is just linear, score based
algorithms,  e.g., quasi-Newton algorithm, are then investigated.
Also based on the disturbance smoothing,  an explicit and simple
expression for the score vector is derived out for both the fixed
effects and variance components involved in MESSM. Based on the
score vector, it is shown   that the maximum likelihood estimation
of MESSM is in fact equivalent to a particular moment estimation.
\par
 As for  state estimation, based on the predicted random effects
 Liu et al (2011) employed Kalman filter to estimate the state. Such
  prediction is  based on the batch data and so it is
 not  a recursive prediction. In many cases, e.g., clinical trial, the
recursive prediction is more meaningful. Furthermore it is known
that the predicting  error of the random effects is rather large if
longitudinal observations are sparse. Ignorance of the predicting
error in this situation will result in a large bias and
underestimate mean squared error of Kalman filter.   In this paper
we propose a algorithm adapted from the algorithm in Liu and West
(2001) to estimate the state which is a recursive method and dose
not require the random effects are known in advance. Thus the
algorithm can apply whether the longitudinal observations are sparse
or not.
\par  In the last the models are further extended to accommodate
  several practical problems, including missing data, non-diagonal
  transition matrix and time-dependent effects et al.
    Simulation examples are carried out which
validate the efficacy of the algorithms of  parameter estimation.
These approaches are applied to a real clinical trial data set and
the results show that though the state estimation is based on the
data only up to the present time point, the resulted  mean squared
errors are comparable to the mean squared error that are resulted
from Kalman filter proposed by Liu et al (2011).
   \par This paper is organized as follows. In section \ref{model formulation},
    the data generating process for generalized MESSM is
   described; In section \ref{model estimation} the algorithms
    for both parameter and state  estimation are detailed;  Several
    further
    extensions
 of the MESSM are considered in section \ref{extensions}.
 In section \ref{numerical studies},
  two numerical examples are investigated to illustrate the efficacy of
 proposed algorithms. Section \ref{conclusions} presents a brief discussion about
 the proposed algorithms.
\section{Model Formulation}\label{model formulation}
Consider a group of  dynamic individuals. For $i$th individual
($i=1,\cdots,m$),  the following linear state space model is
assumed,
\begin{eqnarray}
x_{it}&=&T(\theta_i)x_{i,t-1}+v_{it},\ \ v_{it}\sim N(0,Q),\label{equation 1} \\
y_{it}&=&Z(\theta_i)x_{it}+w_{it},\ \ w_{it}\sim N(0,R),
\label{equation 2}
\end{eqnarray}
where $x_{it}$ and $y_{it}$ are the $p\times 1$ state vector and
$q\times 1$ observation  vector for the $i$th individual at time
$t$;  $v_{it}$ is the $p\times 1$ state disturbance and $w_{it}$ is
the $q\times 1$ observational error, both of which are normally
distributed with mean zero and variance matrix $Q$ and $R$
respectively. The $p\times p$ state transition matrix $T(\theta_i)$
and the $q\times p$ observation matrix $Z(\theta_i)$ are
parameterized with the $r\times 1$ parameter vector $\theta_i$.
\par For $\{v_{it},t=1,2,\cdots\}$,  the following correlation structure are assumed
$$Cov(v_{it},v_{i't'})=\left\{
\begin{array}{lll}
Q(i,i')_{p\times p}&&\mbox{if}\ \  t=t'\\
0 &&\mbox{else}\\
\end{array}
\right.,\label{assumption 2} $$ i.e.,  at the same time point, the
covariance  between the different individuals $i$ and $i'$ is
$Q(i,i')$ and so the individuals in this group are correlated  with
each other. If $i=i'$, then $Q(i,i')=Q$. More complex relationship
also can be possible, see section \ref{general transition matrix}
for another modeling of the relationship among the individuals.  For
$\{w_{it},t=1,2,\cdots\}$, we assume
$$Cov(w_{it},w_{i't'})=\left\{
\begin{array}{lll}
R_{q\times q}&&\mbox{if}\ \ i=i',  t=t'\\
0&&\mbox{else}\ \
\end{array}
\right..\label{assumption 3}$$
 There is another layer of
complexity in model (\ref{equation 1}) $\sim$ (\ref{equation 2}),
i.e., we have to specify the correlation structure for
$\theta_i,(1\leq i \leq n)$, for which  we assume
\begin{eqnarray}\label{theta}
\theta_i&=&\psi_ia+b_i, \ \ b_i\sim N(0,D),
\end{eqnarray}
where $\psi_i$ is  the  exogenous variable representing the
characteristics of the $i$th individuals,   $a$ is the fixed effect
and $b_i$ the random effect. We assume $b_i$'s are independent with
$Cov(b_{i},b_{i'})=D$.  Here an implicit assumption is that the
individual parameter $\theta_i$ is static. Time-dependent $\theta_i$
may be more appropriate in some cases which will be considered in
section \ref{general transition matrix}.
 For the correlation structure among
$v_{it}, w_{it}$ and $\theta_i$, we assume
\begin{eqnarray}
Cov(\theta_{i},v_{i't'})=Cov(\theta_{i},w_{i't'})=
Cov(v_{it},w_{i't'})=0 \label{assumption 4}
\end{eqnarray}
 for $1\leq i \leq m,
1\leq i'\leq m, t\geq 1,t'\geq 1$.
\par
The model given above is  a generalized version of MESSM given in
Liu et al (2011) and Lodewyckx et al (2011), in which  they  assume
that the disturbance $v_{it}$ is independent to $v_{i't}$ for $i\neq
i'$.  Here we assume  there exists static correlation among the
individuals. Another critical assumption in Liu et al (2011) is that
both $T(\theta_i)$ and $Z(\theta_i)$ should be the linear functions
of $\theta_i$. Here this restriction also is not required.
\par The following  notations are adopted in this paper. $\{_m \
a_{ij}\}{_{i=1}^p}{_{j=1}^q} =\{_m \ a_{ij}\}$ denotes a $p\times q$
matrix with elements $a_{ij}$;  $\{_c\ u_i\}_{i=1}^n$ denotes a $n$
dimensional  column vector; $\{_r\ u_i\}_{i=1}^m$ denotes a $n$
dimensional row vector; diagonal matrix is denoted by $\{_d\
a_i\}_{i=1}^n$. All the elements can be replaced by matrixes  which
will result in a block matrix. As for the model (\ref{equation
1})$\sim$(\ref{equation 2}), define $x_t=\{_c\ x_{it}\}_{i=1}^m$,
$\theta=\{_c\ \theta_{i}\}_{i=1}^m$, $\tilde{T}(\theta)=\{_d\
T(\theta_i)\}_{i=1}^m$, $\tilde{Z}(\theta)=\{_d\
Z(\theta_i)\}_{i=1}^m$,
 $v_t=\{_c\ v_{it}\}_{i=1}^m$,
$y_t=\{_c\ y_{it}\}_{i=1}^m$, $w_t=\{_c\ w_{it}\}_{i=1}^m$, and then
the model can be written in matrix form as
\begin{eqnarray}
x_t&=&\tilde{T}(\theta)x_{t-1}+v_t,\label{model1}\\
y_t&=&\tilde{Z}(\theta)x_{t}+w_t.\label{model2}
\end{eqnarray}
Here  $Var(v(t))\triangleq \tilde{Q}={ \{_m\
Q(i,i')\}_{i=1}^m}_{i'=1}^m$, $Var(w_t)\triangleq \tilde{R}=\{_d\
R\}_{i=1}^m$, $Var(\theta)=\{_d\ D\}_{i=1}^m$  and
$Cov(v_t,w_t)=Cov(\theta,w_t)=Cov(\theta,w_t)=0$. Equations
(\ref{equation 1})$\sim$(\ref{model2}) represent the data generating
process. Given the observations up to time $t$,
$y_{1:t}=(y_{11},\cdots,y_{m1},\cdots,y_{1t},\cdots,y_{mt})$, we
will study the following problems: (1)\ How to estimate the
parameters involved in the model,
 including  covariance matrix  $\tilde{Q}$, $\tilde{R}$, $D$ and fixed effects
  $a$. (2)\  How to get the online
estimate of the state  $x_{it}$ for $1\leq i \leq
 m$.  Though These problems had been
 studied in literatures, we will adopt  different
 ways to address these issues which turn out  to be more efficient in most settings.
\section{Model Estimation}\label{model estimation}
 The parameters  involved in MESSM include the fixed effects
 $a$ and those involved in  variance matrixes $ (\tilde{Q}, \tilde{R}, D)$  which is
 denoted by $\delta$. We write $(\tilde{Q}(\delta), \tilde{R}(\delta),
 D(\delta))$ to indicate explicitly  such  dependence of variance matrix on $\delta$.
In this section we consider how to estimate parameter
$\Delta^T\triangleq(a^T,\delta^T)$ and the state $x_t$ based on the
observations $y_{1:T}$. Lodewyckx et al (2011) and Liu et al (2011)
had investigated these questions in details, including EM algorithm
based maximum likelihood estimation and Baysian estimation. While
these approaches are shown to be efficient for the given
illustrations, they  are cumbersome to be carried out. On the other
hand it is also well known that the  rate of convergence for EM
algorithm is linear which is slower than quasi-Newton algorithm. In
the following we will first consider a new version of  EM algorithm
which is simpler than the existed results. Then scores based
algorithm is investigated. Explicit and simple expression for the
score vector is derived out. State estimation also is investigated
using an adapted filter algorithm proposed by Liu and West (2001).
\subsection{Maximizing the  likelihood  via EM algorithm}
\label{EM} For model (\ref{model1})$\sim$(\ref{model2}),  we take
$(\theta^T,x_1^T,\cdots,x_n^T)^T$ as the missing data and
$(\theta^T, x_1^T,\cdots,x_n^T,y_{1:T})^T$ the complete data. Note
$f(\theta,x_{1:T},y_{1:T}|\Delta)=f(\theta|\Delta)f(x_{1:T}|\theta,\Delta)
f(y_{1:T}|\theta,x_{1:T},\Delta)$ in which all the terms
$f(\theta|\Delta)$, $f(x_{1:T}|\theta,\Delta)$ and
$f(y_{1:T}|\theta,x_{1:T},\Delta)$ are normal densities by
assumption. For the sake of simplicity, we let $x_1\sim N(a_1,P_1)$
with known $a_1$ and $P_1$.  Then omitting constants, the log joint
density can be written as
\begin{eqnarray}\label{jointdensity}
&&\log f(\theta,x_{1:T},y_{1:T}|\Delta)=-\frac{m}{2}\log |D(\delta)|
-\frac{T}{2} \log |\tilde{R}(\delta)|-\frac{T}{2}\log
|\tilde{Q}(\delta)|\\
&&-\frac{1}{2}\sum_{i=1}^m {\rm
tr}\left[D(\delta)^{-1}(\theta_i-\Psi_i a)(\theta_i-\Psi_i
a)^T\right]-\frac{1}{2}\sum_{t=1}^T {\rm tr}[\tilde{R}(\delta)^{-1}
\{y_{t}-\tilde{Z}(\theta)x_{t}\}\nonumber\\
&&\times
\{y_{t}-\tilde{Z}(\theta)x_{t}\}^T]-\frac{1}{2}\sum_{t=1}^T {\rm
tr}[\tilde{Q}(\delta)^{-1}\{x_t-\tilde{T}(\theta)x_{t-1}\}
\{x_t-\tilde{T}(\theta)x_{t-1}\}^T]\nonumber
\end{eqnarray}
where for $t=1$,
$\tilde{Q}(\delta)^{-1}\{x_t-\tilde{T}(\theta)x_{t-1}\}
\{x_t-\tilde{T}(\theta)x_{t-1}\}^T$  is explained as
$P_1^{-1}(x_1-a_1)(x_1-a_1)^T$.  Let
$\Delta^{\star}=(a^{\star},\delta^{\star})^T$ denote the value of
$\Delta$ in the $j$th step of  EM algorithm, then
$Q(\Delta,\Delta^{\star})$, the intermediate quantity of EM
algorithm,  is defined as the  expectation of $ \log
f(\theta,x_{1:T},y_{1:T})$ conditional on $\Delta^{\star}$ and the
observations $y_{1:T}$. Let $\tilde{E}(\cdot)$ denote this
conditional expectation and then  with (\ref{jointdensity}) and the
normal assumption in hand, we have
\begin{eqnarray}\label{E-step}
Q(\Delta,\Delta^{\star})&\triangleq&
\tilde{E}[\log f(\theta,x_{1:T},y_{1:T}|\Delta)]
\end{eqnarray}
\begin{eqnarray*}
&=& -\frac{m}{2}\log |D(\delta)| -\frac{T}{2}\log
|\tilde{R}(\delta)|-\frac{T}{2}\log |\tilde{Q}(\delta)|\\ \nonumber
&&-\frac{1}{2}\sum_{i=1}^m {\rm
tr}\left[D(\delta)^{-1}\left\{(\Psi_i(a^{\star}-a)+b_{i|T})
(\Psi_i(a^{\star}-a)+b_{i|T})^T\right.\right.\nonumber\\
&&\left.\left.+{\rm Var}(b_i|y_{1:t},\Delta^{\star}) \right\}\right]
-\frac{1}{2}\sum_{t=1}^T {\rm
tr}\left[\tilde{R}(\delta)^{-1}\left\{w_{t|T}w_{t|T}^T+{\rm
Var}(w_t|y_{1:T},\Delta^{\star})\right\}\right]\nonumber\\
&&-\frac{1}{2}\sum_{t=1}^T {\rm
tr}\left[\tilde{Q}(\delta)^{-1}\left\{v_{t|T}v_{t|T}^T+{\rm
Var}(v_t|y_{1:T},\Delta^{\star})\right\}\right],\nonumber
\end{eqnarray*}
where $ b_{i|T}=\tilde{E}(b_i), w_{t|T}=\tilde{E}(w_{t}),
 v_{t|T}=\tilde{E}(v_t)$. In order to find
 the maximizer of $Q(\Delta, \Delta^{\star})$ with respect to
 $\Delta$, we have to compute these conditional expectations and
variances firstly. Note that
\begin{eqnarray}
b_{i|T}= \tilde{E}(b_{i|T}(\theta)),\quad
w_{t|T}=\tilde{E}(w_{t|T}(\theta)),\quad
 v_{t|T}= \tilde{E}(v_{t|T}(\theta)),\label{expectation}
\end{eqnarray}
where
\begin{eqnarray*}
b_{i|T}(\theta)=E(b_i|y_{1:T},\Delta^{\star},\theta),\
w_{t|T}(\theta)=E(w_t|y_{1:T},\Delta^{\star},\theta),\
v_{t|T}(\theta)=E(v_t|y_{1:T},\Delta^{\star},\theta)
\end{eqnarray*}
 and
\begin{eqnarray}
{\rm Var}(v_t|y_{1:T},\Delta^{\star})&=&\tilde{E}({\rm
Var}(v_t|y_{1:T}, \Delta^{\star},\theta))+{\rm
Var}(v_{t|T}(\theta)|y_{1:T},\Delta^{\star}),\label{variance1}\\
{\rm Var}(w_t|y_{1:T},\Delta^{\star})&=&\tilde{E}({\rm
Var}(w_t|y_{1:T}, \Delta^{\star},\theta))+{\rm
Var}(w_{t|T}(\theta)|y_{1:T},\Delta^{\star}).\label{variance2}
\end{eqnarray}
  For the  smoothed disturbances $w_{t|T}(\theta),
v_{t|T}(\theta)$ and the relevant  variances  we have,
\begin{eqnarray}
w_{t|T}(\theta)=\tilde{R}(\delta^{\star})e_t(\theta),&& {\rm
Var}(w_t|y_{1:T},\Delta^{\star},\theta)=\tilde{R}(\delta^{\star})-
\tilde{R}(\delta^{\star})D_t(\theta)\tilde{R}(\delta^{\star}),\label{recursion1}\\
v_{t|T}(\theta)=\tilde{Q}(\delta^{\star})r_{t-1}(\theta),&& {\rm
Var}(v_t|y_{1:T},\Delta^{\star},\theta)=\tilde{Q}
(\delta^{\star})-\tilde{Q}(\delta^{\star})N_{t-1}(\theta)
\tilde{Q}(\delta^{\star}),
\end{eqnarray}
where the backward recursions for $e_t(\theta)$,  $r_{t}(\theta)$,
$D_t(\theta)$ and $N_t(\theta)$  are given by
\begin{eqnarray}
e_t(\theta)&=&F_t(\theta)^{-1}\nu_t-K_t(\theta)^T r_t(\theta),\label{e and D1}\\
r_{t-1}(\theta)&=&Z(\theta)^T
F_t^{-1}(\theta)\nu_t+L_t(\theta)^Tr_t(\theta),\label{e and D2}\\
D_t(\theta)&=&F_t(\theta)^{-1}+K_t(\theta)^TN_t(\theta)K_t(\theta),\label{r and N1}\\
N_{t-1}(\theta)&=&Z(\theta)^T F_t(\theta)^{-1}Z(\theta)
+L_t(\theta)^TN_t(\theta)L_t(\theta)\label{r and N2}
\end{eqnarray}
for  $t=T,\cdots,1$.  These terms are calculated backwardly with
$r_T=0$ and $N_T=0$. Here $F_t(\theta), K_t(\theta)$ are
respectively the variance matrix of innovation and gain matrix
involved in Kalman filter. The recursions for these matrix   can be
stated as follows,
\begin{eqnarray}
P_{t+1|t}(\theta)=T(\theta)P_{t|t-1}(\theta)L_t^T+\tilde{Q}(\delta^{\star}),&&
F_t(\theta)=Z(\theta)P_{t|t-1}(\theta)Z(\theta)^T+\tilde{R}(\delta^{\star}),\\
K_t(\theta)=T(\theta)P_{t|t-1}(\theta)Z(\theta)^TF_t(\theta)^{-1},&&
L_t(\theta)=T(\theta)-K_t(\theta)Z(\theta).\label{recursion2}
\end{eqnarray}
The recursions  (\ref{recursion1})$\sim$(\ref{recursion2}) can be
found in  Durbin and Koopman (2012).  Combining these recursive
formulas with (\ref{expectation})$\sim$(\ref{variance2}) yields
\begin{eqnarray*}
w_{t|T}w_{t|T}^T&=&\tilde{R}(\delta^{\star})\tilde{E}(e_t(\theta))
\tilde{E}(e_t(\theta))^T \tilde{R}(\delta^{\star}),\label{w}\\
v_{t|T}v_{t|T}^T&=&\tilde{Q}(\delta^{\star})\tilde{E}(r_{t-1}(\theta))
\tilde{E}(r_{t-1}(\theta))^T \tilde{Q}(\delta^{\star}),\label{v}\\
{\rm Var}(w_t|y_{1:T},\Delta^{\star})&=&\tilde{R}(\delta^{\star})-
\tilde{R}(\delta^{\star})\tilde{E}(D_t(\theta))\tilde{R}(\delta^{\star})
+ \tilde{R}(\delta^{\star}){\rm
Var}(e_t(\theta)|y_{1:T},\Delta^{\star})\tilde{R}
(\delta^{\star}),\label{e}\\
{\rm Var}(v_t|y_{1:T},\Delta^{\star})&=&\tilde{Q}(\delta^{\star})-
\tilde{Q}(\delta^{\star})\tilde{E}(N_{t-1}(\theta))\tilde{Q}(\delta^{\star})
+ \tilde{Q}(\delta^{\star}){\rm
Var}(r_{t-1}(\theta)|y_{1:T},\Delta^{\star})\tilde{Q}(\delta^{\star}).\label{r}
\end{eqnarray*}
Substituting these expression    into (\ref{E-step})  we have
\begin{eqnarray}\label{Q function}
Q(\Delta,\Delta^{\star}) &=& -\frac{m}{2}\log |D(\delta)|
-\frac{T}{2}\log |\tilde{R}(\delta)|-\frac{T}{2}\log
|\tilde{Q}(\delta)|\\
 &&-\frac{1}{2}\sum_{i=1}^m {\rm
tr}\left[D(\delta)^{-1}\left\{(\Psi_i(a^{\star}-a)+b_{i|T})
(\Psi_i(a^{\star}-a)+b_{i|T})^T\right.\right.\nonumber\\
&&\left.\left.+{\rm Var}(b_i|y_{1:t},\Delta^{\star})
\right\}\right]\nonumber\\
&& -\frac{1}{2}\sum_{t=1}^T {\rm
tr}\left[\tilde{R}(\delta)^{-1}\left\{\tilde{R}(\delta^{\star})+
\tilde{R}(\delta^{\star})\tilde{E}(e_t^2(\theta)-D_t(\theta))\tilde{R}(\delta^{\star})
\right\}\right]\nonumber\\
&&-\frac{1}{2}\sum_{t=1}^T {\rm
tr}\left[\tilde{Q}(\delta)^{-1}\left\{\tilde{Q}(\delta^{\star})+
\tilde{Q}(\delta^{\star})\tilde{E}(r_{t-1}^2(\theta)-N_{t-1}(\theta))\tilde{Q}(\delta^{\star})
\right\}\right].\nonumber
\end{eqnarray}
\par Now we have obtained  the  expression for the intermediate quantity of EM algorithm.
Except conditional expectations and variances, all the quantities
involved  can be easily computed by Kalman filter. These conditional
expectations and variances include $b_{i|T}$,
$\tilde{E}D_t(\theta)$, $\tilde{E}N_{t}(\theta)$,
$\tilde{E}\{e_t(\theta)e_t(\theta)^T\}$,  $\tilde{E}\{r_{t-1}
(\theta)r_{t-1}(\theta)^T\}$ and $Var(b_i|y_{1:t},\Delta^{\star})$.
Here we adopt the Monte Carlo method to approximate the expectations
and variances. Specifically given the random samples
$\{\theta^{(j)}_t,j=1,\cdots,M\}$ from the
 posterior $f(\theta|y_{1:T}, \Delta^{\star})$, all the population
expectation is approximated by the sample expectation. For example
we approximate  $b_{i|T}$ by $\frac{1}{M} \sum_{j=1}^M
\theta^{(j)}_i-\Psi_i a^{\star}$. The same approximation applies to
other expectations and variances. As for the sampling from the
posterior $f(\theta|y_{1:T},\Delta^{\star})$,  the random-walk
Metropolis algorithm   is employed in this paper to generate the
samples. Certainly it is also  possible to  use other sampling
scheme such as importance sampling to generate the random samples
from $f(\theta|y_{1:T},\Delta^{\star})$. In our finite experiences
MCMC algorithm is superior to the importance sampling in present
situations. It is meaningful to compare the proposed EM algorithm
with that in Liu et al (2011). Recall the EM algorithm in Liu et al
(2011) have to sample both $x_t$ and $\theta$ from the joint
distribution $f(\theta,x_t|y_{1:T},\Delta^{\star})$ where Gibbs
sampler was proposed to implement the sampling  in their study. Here
only the random samples of $\theta$ from
$f(\theta|y_{1:T},\Delta^{\star})$ are needed for running the EM
algorithm and thus the proposed EM algorithm can be seen as a
Rao-Blackwellized version of that in Liu et al (2011). Note the
dimension of $x_t$ increases as the number of the individuals
increases and  consequently a faster and more stable convergence
rate of the proposed algorithm can be expected especially when the
number of the correlated individuals is large.
\par For the purpose of illustration consider the following autoregressive
plus noise model,
\begin{eqnarray}\label{autoregression1}
y_{it}=x_{it}+w_{it},\  w_{it}\sim {\rm i.i.d.}\  N(0,\delta_1),\
x_{it}=\theta_ix_{i,t-1}+v_{it},\ v_{it}\sim {\rm i.i.d.}\
N(0,\delta_2)
\end{eqnarray}
where \begin{eqnarray}\label{autoregression2}
\theta_i&=&\mu_{\theta}+b_i,\ b_i\sim {\rm i.i.d.}\ N(0,\delta_3),
i=1,\cdots,m.
\end{eqnarray}
Here we assume all the individuals in this group are independent
with each other. This model can be rewritten in the matrix form as
\begin{eqnarray*}
y_t=\tilde{Z}(\theta)x_t+w_t,w_t\sim N_m(0,\delta_1 I_m), &&
x_t=\tilde{T}(\theta)x_{t-1}+v_t, v_t\sim N_m(0,\delta_2 I_m)
\end{eqnarray*}
with $y_t=(y_{1t},\cdots,y_{mt})^T$, $x_t=(x_{1t},\cdots,x_{mt})^T$,
 $w_t=(w_{1t},\cdots,w_{mt})^T$,  $v_t=(v_{1t},\cdots,v_{mt})^T$,
 $\theta=(\theta_1,\cdots,\theta_m)^T$, $\delta=(\delta_1,\delta_2,\delta_3)^T$,
 $\tilde{Z}(\theta)=I_m$, $\tilde{T}(\theta)={\rm diag}(\theta_1,\cdots,\theta_m)$,
$\tilde{R}(\delta)=\delta_1I_m$, $\tilde{Q}=\delta_2I_m$,
$\Psi_i=1$, $D(\delta)=\delta_3$.  From (\ref{Q function}) we get
the following recursive formulas,
\begin{eqnarray}\label{recursive1}
\hat{\mu}_{\theta}&=&\frac{1}{m}\sum_{i=1}^m
\tilde{E}(\theta_i),\quad \hat{\delta}_3=\frac{1}{m}\sum_{i=1}^m
[\hat{\mu}_{\theta}-\tilde{E}(\theta_i)]^2,\nonumber\\
\hat{\delta}_1&=&\frac{1}{Tm}\sum_{i=1}^m \sum_{t=1}^T
[\delta_1^{\star}+{\delta_1^{\star}}^2\tilde{E}\{e_{it}^2(\theta)-D_{it}(\theta)\}],\\
\hat{\delta}_2&=&\frac{1}{Tm}\sum_{i=1}^m \sum_{t=1}^T
[\delta_2^{\star}+{\delta_2^{\star}}^2\tilde{E}\{r_{it-1}^2(\theta)
-N_{it-1}(\theta)\}].\nonumber
\end{eqnarray}
After getting $(\hat{\mu}_{\theta},\hat{\delta}_1, \hat{\delta}_2,
\hat{\delta}_3 )^T$ from (\ref{recursive1}), we take it as the new
$\Delta^{\star}$ and use it to  compute the next maximizer of
$Q(\Delta,\Delta^{\star})$ until the convergence is achieved. The
convergent point is defined as the  estimator of $\Delta$.
\par The second illustration we  consider is the damped local linear
model which can be expressed as
\begin{eqnarray}\label{local linear1}
y_{it}&=&z_{it}+\epsilon_{it},\quad
z_{it}=z_{i(t-1)}+u_{it}+\eta_{it},\quad u_{it}=\theta_i
u_{i(t-1)}+\tau_{it},
\end{eqnarray}
with $\theta_i=\mu_{\theta}+b_i$ and
\begin{eqnarray}\label{local linear2}
\epsilon_{it}\sim {\rm i.i.d.} N(0,\delta_1),\quad\eta_{it}\sim {\rm
i.i.d.} N(0,\delta_2),\quad\tau_{it}\sim {\rm i.i.d.} N(0,\delta_3),
\quad b_i\sim {\rm i.i.d.} N(0,\delta_4).
\end{eqnarray}
We also assume that the individuals in the group are independent
with each other. Defining the state variable as
$x_{it}=(z_{it},u_{it})^T$, then the damped local linear model can
be rewritten as the state space model (\ref{equation
1})$\sim$(\ref{equation 2}) with
$$T=(1,0),\ Z
=\left(
\begin{array}{cc}
1&1\\
0&\theta_i\\
\end{array}
\right),\ Q=\left(
\begin{array}{cc}
\delta_2&0\\
0&\delta_3\\
\end{array}
\right),\ R=\delta_1.$$ Here the unknown parameters include
$\Delta\triangleq(\mu_{\theta},\delta_1,\delta_2,\delta_3,\delta_4)^T$.
Let $$r_{it}\triangleq(r_{it}^{(z)},  r_{it}^{(u)})^T,\quad N_{it}
\triangleq\left(
\begin{array}{cc}
N_{it}^{(zz)}&N_{it}^{(zu)}\\
N_{it}^{(zu)}&N_{it}^{(uu)}\\
\end{array}
\right),$$ then from (\ref{Q function}),  the recursive formula of
EM algorithm turns out to be
\begin{eqnarray*}
\hat{\mu}_{\theta}&=&\frac{1}{m}\sum_{i=1}^m
\tilde{E}(\theta_i),\quad \hat{\delta}_4=\frac{1}{m}\sum_{i=1}^m
[\hat{\mu}_{\theta}-\tilde{E}(\theta_i)]^2,\\
\hat{\delta}_1&=&\frac{1}{Tm}\sum_{i=1}^m \sum_{t=1}^T
[\delta_1^{\star}+{\delta_1^{\star}}^2\tilde{E}\{e_{it}^2(\theta)-D_{it}(\theta)\}],\\
\hat{\delta}_2&=&\frac{1}{Tm}\sum_{i=1}^m \sum_{t=1}^T
[\delta_2^{\star}+{\delta_2^{\star}}^2\tilde{E}\{(r_{it-1}^{(z)})^2(\theta)-
N_{it-1}^{(zz)}(\theta)\}],\\
\hat{\delta}_3&=&\frac{1}{Tm}\sum_{i=1}^m \sum_{t=1}^T
[\delta_3^{\star}+{\delta_3^{\star}}^2\tilde{E}\{(r_{it-1}^{(u)})^2(\theta)-
N_{it-1}^{(uu)}(\theta)\}].
\end{eqnarray*}
\subsection{Maximizing the  likelihood  via score based  algorithms}
In this section we consider the score-based algorithms which include
quasi-Newton algorithm, steepest ascent algorithm et al. The core of
such  algorithms is how to compute the  score vector. Here the
likelihood
 $ L(\Delta|y_{1:t})$ is a complex function of $\Delta$ and the
 direct computation of score is difficult both analytically and
 numerically. We consider the following transformation of
 $L(\Delta|y_{1:t})$,
 \begin{eqnarray}\label{likelihood}
\log L(\Delta|y_{1:T})&=&\log f(\theta,x_{1:T},y_{1:T}|\Delta)-\log
f(\theta,x_{1:T}|y_{1:T}, \Delta)
 \end{eqnarray}
 Recall in section \ref{model estimation}
  $f(\theta,x_{1:T},y_{1:T}|\Delta)$ denotes the joint
 distribution of $(\theta,x_{1:T},y_{1:T})$ conditional on $\Delta$
 and $\tilde{E}(\cdot)$  the conditional expectation
    $E(\cdot|y_{1:T},\Delta^{\star})$. In present situation we let $\Delta^{\star}$
    denote the present value of $\Delta$ in quasi-Newton
    algorithm. Then taking $\tilde{E}$ of both sides of (\ref{likelihood}) yields
 \begin{eqnarray}
 \log
 L(\Delta|y_{1:T})&=&\tilde{E}\left[\log f(\theta,x_{1:T},y_{1:T}|\Delta)\right]-
 \tilde{E}\left[\log f(\theta,x_{1:T}|y_{1:T},\Delta)\right].
 \end{eqnarray}
 Under the assumption that the exchange of integration  and differentiation
   is legitimate it can be shown that
   \begin{eqnarray}
\tilde{E}\left[\left.\frac{\partial  \log
f(\theta,x_{1:T}|y_{1:T},\Delta)}{\partial
\Delta}\right|_{\Delta=\Delta^{\star}}\right]=0,
   \end{eqnarray}
   Consequently we have
\begin{eqnarray}\label{score1}
\left.\frac{\partial \log L(\Delta | y_{1:t})}{\partial
\Delta}\right|_{\Delta=\Delta^{\star}}&=&\frac{\partial }{\partial
\Delta}\left.\tilde{E}\left[ \log f(\theta,x_{1:T},y_{1:T}|\Delta)
\right]\right|_{\Delta=\Delta^{\star}}.
\end{eqnarray}
 Note the expectation in the right-hand side has the same form as
 the intermediate quantity of EM algorithm in the previous
 section. And so
  substituting (\ref{Q function}) into (\ref{score1})  we get
\begin{eqnarray}\label{score2}
\left.\frac{\partial \log L(\Delta | y_{1:T})} {\partial
a}\right|_{\Delta=\Delta^{\star}}&=&\sum_{i=1}^m
\psi_i^TD(\delta^{\star})^{-1}b_{i|T},\\
 \left.\frac{\partial \log L(\Delta |
y_{1:T})}{\partial \delta_j}\right |_{\Delta=\Delta^{\star}}&=&
-\frac{1}{2}\sum_{i=1}^m {\rm tr } \left[
D(\delta^{\star})^{-1}\frac{\partial D(\delta^{\star})}{\partial
\delta_j}\right.\label{score3}\\
&&-D(\delta^{\star})^{-1}\left\{b_{i|T}b_{i|T}^T+Var(b_i|y_{1:T},
\Delta^{\star})\right\} \left.D(\delta^{\star})^{-1}\frac{\partial
D(\delta^{\star})}{\partial \delta_j}\right]\nonumber\\
&&+\frac{1}{2}\sum_{t=1}^T {\rm tr}\left[\tilde{E}\left\{e_t(\theta)
e_t(\theta)^T-D_t(\theta)\right\}\frac{\partial
\tilde{R}(\delta^{\star})}{\partial \delta_j}\right]\nonumber\\
&&+\frac{1}{2}\sum_{t=1}^T {\rm tr}\left[\tilde{E}\left\{r_{t-1}
(\theta)r_{t-1}(\theta)^T-N_{t-1}(\theta)\right\}\frac{\partial
\tilde{Q}(\delta^{\star})}{\partial \delta_j}\right]\nonumber
\end{eqnarray}
 \par
Inspection of the score vector (\ref{score2})$\sim$(\ref{score3})
shows that in order to evaluate  the score vector in present value
$\Delta^{\star}$, we need (1) a single pass of Kalman filter and
smoother, (2) to run a MCMC algorithm to get the random samples
$\theta^{(j)}_{t} (j=1,\cdots,M)$ from
$f(\theta|y_{1:t},\Delta^{\star})$. These calculation can be carried
out readily. It is interesting  to compare this result with the
existed results for the fixed-effects state space models. Engle and
Watson (1981) had constructed a set of filter for computing the
score vector analytically. However, as pointed out by Koopman and
Shephard (1992), this approach is cumbersome, difficult to program
and typically much more expensive to use than numerically
differentiating the likelihood. Koopman and Shephard (1992) and
Koopman (1993) also obtained an analytical expression for the score
vector. But those expressions are only feasible for the variance
components and  the scores for the parameters in observational
matrix and state transition matrix should be computed by numerically
differentiating. On the contrary the exact expressions of  score
vectors given in (\ref{score2})$\sim$(\ref{score3}) not only can be
used to compute the scores for variance components  but also can be
used to compute the scores for fixed effects straightforwardly.
\par As an illustration consider the autoregressive  plus noise model  given by
(\ref{autoregression1})$\sim$(\ref{autoregression2}).  The scores
defined in (\ref{score2})$\sim$(\ref{score3}) can be
 shown to be
\begin{eqnarray}
\frac{\partial \log L(\Delta^{\star}|y_{1:T})}{\partial
\mu_{\theta}}&=&\sum_{i=1}^m \frac{E(\theta_i|y_{1:T},
\Delta^{\star})-\mu_{\theta}^{\star}}{\delta_3^{\star}},\label{exam1_score1}\\
\frac{\partial \log L(\Delta^{\star}|y_{1:T})}{\partial
\delta_1}&=&\frac{1}{2}\sum_{t=1}^T \sum_{i=1}^m
\left[\tilde{E}\left\{e_{it}^2(\theta_i)-D_{it}(\theta_i)\right\}\right],
\label{exam1_score2}\\
\frac{\partial \log L(\Delta^{\star}|y_{1:T})}{\partial
\delta_2}&=&\frac{1}{2}\sum_{t=1}^T \sum_{i=1}^m \left[\tilde{E}
\left\{r_{it-1}^2(\theta_i)-N_{it-1}(\theta_i)\right\}\right], \label{exam1_score3}\\
\frac{\partial \log L(\Delta^{\star}|y_{1:T})}{\partial
\delta_3}&=&-\frac{1}{2}\sum_{i=1}^m \frac{\delta_3^{\star}
-b_{i|T}^2-Var(b_i|y_{1:T},\Delta^{\star})}{\delta_3^{\star
2}}.\label{exam1_score4}
\end{eqnarray}
Here $e_{it}(\theta_i), r_{it}(\theta_i), D_{it}(\theta_i)$ and
$N_{it}(\theta_i)$ have been   defined in (\ref{e and
D1})$\sim$(\ref{r and N2}) which  correspond  to the $i$th
individual. If we denote the MLE of $\Delta$ by
$\hat{\Delta}=(\hat{\mu}_{\theta},\hat{\delta}_1,\hat{\delta}_2,\hat{\delta}_3)^T$,
then  by equating these scores at $\hat{\Delta}$ to zero we have
\begin{eqnarray}
&&\hat{\mu}_{\theta}=\frac{1}{m}\sum_{i=1}^m
E(\theta_i|y_{1:T},\hat{\Delta}),\
\hat{\delta}_3=\frac{1}{m}\sum_{i=1}^m\left(
b_{i|T}^2+Var(b_i|y_{1:T},\hat{\Delta})\right),\label{exam1_1}\\
&&\hspace*{30pt}\sum_{i=1}^m\sum_{t=1}^T
\left[\tilde{E}\{e_{it}^2(\theta_i)\}\right]=
\sum_{i=1}^m\sum_{t=1}^T
\left[\tilde{E}D_{it}(\theta_i)\right],\label{exam1_2}\\
&&\hspace*{30pt}\sum_{i=1}^m\sum_{t=1}^T
\left[\tilde{E}\{r_{it-1}^2(\theta_i)\}\right]= \sum_{i=1}^m
\sum_{t=1}^T
\left[\tilde{E}N_{it-1}(\theta_i)\right].\label{exam1_3}
\end{eqnarray}
  Equations (\ref{exam1_1}) says  that $\hat{\mu}_{\theta}$ is  the sample
 mean of posterior mean  $\tilde{E}(\theta_i)$ at $\Delta=\hat{\Delta}$;
 As for the second term in  (\ref{exam1_1}), note at the true parameter $\Delta_0$,
 $$E\{b_{i|T}^2+Var(b_i|y_{1:T},\Delta_0)\}=Var(E(b_i|y_{1:T},\Delta_0))
 +EVar(b_i|y_{1:T},\Delta_0),$$
  where the right hand side is just equal to $\delta_3$ and
   so $\hat{\delta}_3$ can also be seen as a moment estimator.
  As for equation (\ref{exam1_2})  and (\ref{exam1_3}), it can be easily
 checked
 that for given $\theta \in \Theta$
\begin{eqnarray}
E\{e_{it}^2(\theta)| \hat{\Delta},\theta\}=D_{it}(\theta),\quad
E\{r_{it}^2(\theta)| \hat{\Delta},\theta\}=N_{it}(\theta),
\end{eqnarray}
i.e.,  (\ref{exam1_2}) and (\ref{exam1_3}) are the moment equation
for estimating $\delta_1$ and $\delta_2$. Consequently
$\hat{\Delta}$ can been regarded as a moment estimator.
\par
As another illustration consider the damped local linear model
defined  by (\ref{local linear1})$\sim$ (\ref{local linear2}). The
score vectors can also be obtained by formulas
(\ref{score2})$\sim$(\ref{score3}). In fact it turns out the scores
with respect to   $\mu_{\theta}$, $\delta_1$ and $\delta_4$ have the
same form as the  scores given in (\ref{exam1_score1}),
(\ref{exam1_score2}) and (\ref{exam1_score4}) respectively. As for
$\delta_2$ and $\delta_3$  we have
\begin{eqnarray}
\frac{\partial \log L(\Delta^{\star}|y_{1:T})}{\partial
\delta_2}&=&\frac{1}{2}\sum_{t=1}^T \sum_{i=1}^m \left[\tilde{E}
\left\{(r_{it-1}^{(z)})^2(\theta_i)-N_{it-1}^{(zz)}
(\theta_i)\right\}\right], \label{exam2_score1}\\
\frac{\partial \log L(\Delta^{\star}|y_{1:T})}{\partial
\delta_3}&=&\frac{1}{2}\sum_{t=1}^T \sum_{i=1}^m \left[\tilde{E}
\left\{(r_{it-1}^{(u)})^2(\theta_i)-N_{it-1}^{(uu)}
(\theta_i)\right\}\right]. \label{exam2_score2}
\end{eqnarray}
Here $r_{it}^{(z)}$,  $r_{it}^{(u)}$, $N_{it}^{(zz)}$ and
$N_{it}^{(zu)}$ have been defined in section \ref{EM}.
\par From  these two illustrations  it can be seen that for i.i.d.
individuals, the maximum likelihood estimation  of MESSM  is
equivalent to the  moment estimation. For the general cases where
the individuals may be correlated, this conclusion also holds but
more complex moment equations are needed in those situations.

\subsection{State  estimation}\label{state estimation}
In this section we discuss the  algorithms for  state estimation of
MESSM under the assumption that the true parameter $\Delta_0$  is
known. If the random effects $b_i$'s are also assumed to be known,
then  Kalman filter can yields the optimal state estimator. Just as
mentioned  in section \ref{introduction}, it is unappropriate to
assume $b_i$'s being known in the setting of sparse longitudinal
data and consequently Kalman filter should not be applied directly.
   \par  One way out is to  define the random effects as the new state variables, then
      MESSM turns out to be a
   nonlinear state space model. Consequently
for the state  filter,  we can employ the usual nonlinear filter or
Monte Carlo filter  to estimate the state.
  Though being straightforward, this approach is thought to be suboptimal because it
   does not utilize  the structure information contained in MESSM
    (\ref{equation 1})$\sim$(\ref{equation 2}) in an  efficient
      way.\par Note that  given the random effects,
        MESSM is a conditional  linear state space model and so the
        mixture Kalman filter proposed in Chen and Liu (2001)
         seems to be a good candidate for state
        estimation. However because the parameter $\theta$  is
static in present settings, the re-sampling step  in mixture Kalman
fitler will make the sample $\left\{\theta_{t}^{(1)}, \cdots,
\theta_{t}^{(M)}\right \}$ at time $t$  being a sub-sample of the
sample $\{\theta_{t-1}^{(1)}, \cdots, \theta_{t-1}^{(M)} \}$ at time
$t-1$. This will make $\{\theta_{t}^{(1)}, \cdots, \theta_{t}^{(M)}
\}$ a poor representative of the posterior $f(\theta|y_{1:t})$ as
time $t$ passes. In order to get an improved representative of
$f(\theta|y_{1:t})$,  in the following we will  present another
algorithm which can overcome the problem of particle degeneracy
 and usually has a better performance in the aspect of representation of
$f(\theta|y_{1:t})$ than usual mixture Kalman filter. This filter
algorithm is adapted from the work in Liu and West (2001). The idea
is to approximate the posterior distribution $f(\theta|y_{1:t})$
sequentially  by a proper mixture of normal distribution. Then the
problem of sampling from the complex posterior $f(\theta|y_{1:t})$
becomes a problem of sampling from a mixture distribution, which can
be carried out straightforwardly. Specifically at time $t$ we assume
the following approximation is appropriate
\begin{eqnarray}
f(\theta|y_{1:t})\approx \sum_{j=1}^M w_{t}^{(j)}N(m_t^{(j)}, h^2
V_t)
\end{eqnarray}
for some proper $w_t^{(j)}$, $m_t^{(j)}$ and $V_t$. The choices of
$w_t^{(j)}$, $m_t^{(j)}$ and $V_t$ depend on the last particles
$\{\theta_{t-1}^{(1)}, \cdots, \theta_{t-1}^{(M)} \}$ and the
present observation $y_t$. The smoothing parameter $h$ controls the
overall scale. We denote the Kalman filter at time $t\geq 1$
corresponding to $\theta_{t}^{(j)}$ by
$KF_{t}^{(j)}=\left(x_{t|t}^{(j)},P_{t|t}^{(j)},x_{t+1|t}^{(j)},P_{t+1|t}^{(j)}
\right)$ where $x_{t|t}^{(j)}$ denotes the filter estimator of $x_t$
with variance $P_{t|t}^{(j)}$; $x_{t+1|t}^{(j)}$ denotes the
one-step-ahead predictor of $x_{t+1}$ with variance
$P_{t+1|t}^{(j)}$.  The filter algorithm can then be stated as
follows.
\par
 Suppose the Monte Carlo sample
$\theta_{t-1}^{(j)}$ and
 weights $w_{t-1}^{(j)}$ ($j=1,\cdots,M$), representing the
 posterior $f(\theta|y_{1:t-1})$,
   are available. Also the Kalman filter $KF_{t-1}^{(j)}$ has been derived out.
   $\bar{\theta}_{t-1}$ and $V_{t-1}$ denote the weighted
    sample mean and variance  of the
   particles $\{\theta_{t-1}^{(1)}, \cdots, \theta_{t-1}^{(M)} \}$
    respectively.
    Then at time $t$ when the observation $y_t$ is brought in,
    \par $\bullet$ For each $j=1,\cdots,M$, compute
    $ m_{t-1}^{(j)}=a\theta_{t-1}^{(j)}+(1-a)\bar{\theta}_{t-1}$ where $a=\sqrt{1-h^2}$.
    \par $\bullet$ Sample an auxiliary integer variable from set $\{1,\cdots,
    M\}$ with probabilities proportional to $z_t^{(j)}\propto w_{t-1}^{(j)}
    f(y_t|x_{t|t-1}^{(j)}, \theta_{t-1}^{(j)})$, which is referred as $k$.
    \par $\bullet$ Sample a new parameter vector $\theta_t^{(k)}$
    from the $k$th normal component of the kernel density, i.e.,
$\theta_t^{(k)}\sim N(m_{t-1}^{(k)}, h^2V_{t-1})$.  \par $\bullet$
For $\theta_t^{(k)}$, compute $KF_t^{(k)}$ and evaluate the
corresponding weight $$w_{t}^{(k)}=\frac{f(y_{t}|x_{t|t}^{(k)},
\theta_t^{(k)})}{ f(y_{t}|x_{t|t-1}^{(k)}, m_{t-1}^{(k)})}$$. \par
$\bullet$ Repeat step (2)-(4) a large number of times to produce a
final posterior approximation $\theta_t^{(k)}$ and Kalman filter
$KF_t^{(k)}$ both of which are associated  with weights $w_t^{(k)}$.
\par We call the algorithm above mixture Kalman filter with kernel smoothing
(MKF-KS).  Historically using kernel smoothing of density to
approximate the posterior distribution of dynamic system originated
from West (1993a,1993b). MKF-KS assumes that the posterior can be
well approximated by a mixture of normal distribution which in many
cases is a reasonable assumption. More important is that  MKF-KS can
solves the  problem of particle degeneration satisfyingly in most
settings.  From the Example 2  in section \ref{numerical studies} it
can be seen MKF-KS does have a good  performance. Therefore we
recommend to use MKF-KS to estimate state for MESSM when the
observations are sparse.\par In additional to state estimation,
MKF-KS can also be used as a basis to estimate the observed
information matrix whose inverse  usually is taken as the estimate
of the variance matrix of the maximum likelihood estimator in
literatures. Poyiadjis et al (2011) is the first to use the particle
filter to approximate the observed information matrix. Nemeth et al
(2013) improved the efficiency of such algorithms by using the  idea
of kernel smoothing of Liu and West (2001). The details of this
algorithm will be omitted for brevity, for further details see
Nemeth et al (2013). In the section \ref{numerical studies}, we will
combine MKF-KS with the algorithms 3 in Nemeth et al (2013) to
estimate the observed information matrix.
\section{Extensions}\label{extensions}
\subsection{Incomplete observations}\label{incomplete observations}
In  previous sections, we have assumed  all the individuals can be
observed  at all the time points. For longitudinal data  however
such assumption  does not hold in many settings and the observations
for some or even all  of individuals may be missing at given time
point. In this section we show that the mixed-effects state space
model can be easily adapted to accommodate such situations.
\par
Assume first the observations for all of the individuals are missing
at time $t$ for $\tau\leq t \leq  \tau^{\star}-1$.  As for the EM
algorithm in section \ref{model estimation}, the intermediate
quantity now is given by (\ref{Q function}) minus the following
terms,
\begin{eqnarray}\label{missing Q}
&&\quad \quad\quad \quad -\frac{\tau^{\star}-\tau}{2}\log
|\tilde{R}(\delta)|-\frac{\tau^{\star}-\tau}{2}\log
|\tilde{Q}(\delta)|\nonumber\\
&&-\frac{1}{2}\sum_{t=\tau}^{\tau^{\star}-1} {\rm
tr}\left[\tilde{R}(\delta)^{-1}\left\{\tilde{R}(\delta^{\star})+
\tilde{R}(\delta^{\star})\tilde{E}(e_t^2(\theta)-D_t(\theta))\tilde{R}(\delta^{\star})
\right\}\right]
\end{eqnarray}
\begin{eqnarray}
&&-\frac{1}{2}\sum_{t=\tau}^{\tau^{\star}-1} {\rm
tr}\left[\tilde{Q}(\delta)^{-1}\left\{\tilde{Q}(\delta^{\star})+
\tilde{Q}(\delta^{\star})\tilde{E}(r_{t-1}^2(\theta)-N_{t-1}(\theta))\tilde{Q}(\delta^{\star})
\right\}\right].\nonumber
\end{eqnarray}
Note here $\tilde{E}(\cdot)$ is interpreted as
$\tilde{E}(\cdot)=E(\cdot|y_{1:\tau-1,\tau^{\star}:T},\Delta^{\star})$.
As for the quasi-Newton algorithm in section \ref{model estimation},
the equation (\ref{score1} ) still holds in the present situation
with the new interpretation of $\tilde{E}(\cdot)$. It can be shown
straightforwardly  that the scores with respect to fixed effects
are the same as that given in (\ref{score2}) while the scores with
respect to variance components are just that given in (\ref{score3})
minus the following terms,
\begin{eqnarray*}
\sum_{t=\tau}^{\tau^{\star}-1} {\rm
tr}\left[\tilde{E}\left\{e_t(\theta)
e_t(\theta)^T-D_t(\theta)\right\}\frac{\partial
\tilde{R}(\delta^{\star})}{\partial
\delta_j}+\tilde{E}\left\{r_{t-1}(\theta)
r_{t-1}(\theta)^T-N_{t-1}(\theta)\right\}\frac{\partial
\tilde{Q}(\delta^{\star})}{\partial \delta_j}\right].
\end{eqnarray*}
\par As for state estimation, the only changes occurs when $\tau\leq t
\leq \tau^{\star}-1$.  Given $\theta^{(j)}$ with $1\leq j \leq M$,
the Kalman filter involved in MKF-KS at time $\tau\leq t \leq
\tau^{\star}-1$
 can be stated as
 \begin{eqnarray}
&&\hspace*{40pt}x_{t|t}^{(j)}=x_{t|t-1}^{(j)},\quad
P_{t|t}^{(j)}=P_{t|t-1}^{(j)},\nonumber\\
&&x_{t+1|t}^{(j)}=T(\theta^{(j)})x_{t|t}^{(j)},\quad
P_{t+1|t}^{(j)}=T(\theta^{(j)})P_{t|t}^{(j)}T(\theta^{(j)})^T+\tilde{Q}.\nonumber
\end{eqnarray}
 While for  weights involved in MKF-KS,
we only need to modify the weight in the second step in MKF-KS from
$w_{t-1}f(y_t|x_{t|t-1}^{(j)},\theta_{t-1}^{(j)})$ to $w_{t-1}$. The
weight in the fourth step will be unchanged.
 \par Another type  of the missing data  is that
 only some of the individuals
 are not observed at given time point. In order to accommodate such case, we
 only need to allow the  observation matrix $\tilde{Z}(\theta)$
being time-dependent. Now model
 (\ref{model1})$\sim$(\ref{model2}) becomes
\begin{eqnarray}
x_t&=&\tilde{T}(\theta)x_{t-1}+v_t,\label{model3}\\
y_t&=&\tilde{Z}_t(\theta)x_{t}+w_t.\label{model4}
\end{eqnarray}
(\ref{model3})$\sim$(\ref{model4}) allow $\tilde{Z}_t(\theta)$ can
possess different dimension at differen time point and thus can
accommodate this type of missing data. The algorithms for parameter
and state estimation given in section \ref{model estimation}   can
be extended straightforwardly to accommodate this more general
model. Example 2 in the next section involves a real data set  which
contains both types of missing data.
\subsection{General transition matrix}\label{general transition
matrix} In section \ref{model formulation}, we have assumed the
individuals in the group can be correlated, i.e., the covariance
matrix $Q(i,i')$ may be a non-diagonal matrix. In addition to
allowing the non-diagonal covariance matrix, the correlation within
the group can also be modeled by adopting a different form of
$\tilde{F}(\theta)$, the state transition matrix.  In section
\ref{model formulation}, we have assumed  $\tilde{F}(\theta)$ is a
diagonal matrix, i.e., $\tilde{F}(\theta)=\{_d\
F(\theta_i)\}_{i=1}^m$. It can be seen that the algorithms of  the
parameter  and state estimation in the previous sections can apply
regardless of $\tilde{F}(\theta)$ being a diagonal matrix or not.
Non-diagonal transition matrix  can occur in many different
situations. Consider the following target tracking model,
\begin{eqnarray}
d\dot{S}_{it}&=&\{-\alpha_i[S_{it}-h(S_t)]-\gamma_i
\dot{S}_{it}-\beta_i[\dot{S}_{it}-g(\dot{S}_t)]\}dt+dW_{it}+dB_t,\label{ground
tracking}
\end{eqnarray}
where $S_{it}=(S_{it}^{(x)},S_{it}^{(y)})^T$ denotes the position of
target $i$ at time $t$;
$\dot{S}_{it}=(\dot{S}_{it}^{(x)},\dot{S}_{it}^{(y)})^T$ denotes the
velocity of target $i$ at time $t$; $h(S_t)=\frac{1}{N}\sum_{i=1}^m
S_{it}$ and $g(\dot{S}_t)=\frac{1}{m}\sum_{i=1}^m \dot{S}_{it}$
denotes the average position and velocity at time $t$.  $B_t$ is a
2-dimensional Brownian motion common to all targets; $W_{it}$ is
another 2-dimensional  Brownian motion assumed to be independently
generated for each target $i$ in the group; $\alpha_i$ denotes the
rate at which $S_{it}$ restores to the average position $h(S_t)$;
$\beta_i$ denotes the rate at which $\dot{S}_{itt}$ restores to the
average velocity $g(\dot{S}_t)$; $\gamma_i$ denotes the rate at
which $\dot{S}_{it}$ restores to zero. Model (\ref{ground tracking})
is the fundamental model for the group tracking problem. In present
literatures, e.g., Khan et al (2005), Pang et al (2008, 2011), three
restoring parameters $\alpha_i,\beta_i, \gamma_i$ are assumed to be
identical across different individuals, i.e.,
$\alpha_1=\cdots=\alpha_m$, $\beta_1=\cdots=\beta_m$,
$\gamma_i=\cdots=\gamma_m$.  Here  with MESSM in hand we can   relax
this restriction and allow different restoring parameters for
different individuals which is  more reasonable in most situations.
Let $\theta_i=(\alpha_i,\beta_i,\gamma_i)$ and
$\theta^T=(\theta_1^T,\cdots,\theta_m^T)$.   For $i=1,\cdots,m$, let
$$A_{i2}=\left(
\begin{array}{cc}
0&1\\
-\alpha_i+\frac{\alpha_i}{m}&-\beta_i-\gamma_i+\frac{\beta_i}{m}
\end{array}\right),\quad
A_{i4}=\left(
\begin{array}{cc}
0&0\\
\frac{\alpha_i}{m}& \frac{\beta_i}{m}
\end{array}\right),
$$
and $A_{i1}=\{_d\ A_{i2},A_{i2}\}$, $A_{i3}=\{_d\ A_{i4}, A_{i4}\}$,
$$A(\theta)=\left(
\begin{array}{cccc}
A_{11}&A_{13}&\cdots&A_{13}\\
A_{23}&A_{21}&\cdots&A_{23}\\
\cdot&\cdot&\cdots&\cdot\\
\cdot&\cdot&\cdots&\cdot\\
A_{m3}&A_{m3}&\cdots&A_{m1}
\end{array}
\right)_{4m\times 4m}. \label{assumption 1}$$ Defining  the
non-diagonal matrix $T(\theta)=\exp (A(\theta)\tau)$ where $\tau$ is
the time between successive observations , then we have the
following discretized version of model (\ref{ground tracking}) for
$m$ targets,
\begin{eqnarray}\label{state equation}
x_t=T(\theta)x_{t-1}+v_t,
\end{eqnarray}
where $x_t=(S_{1t}^{(x)},\dot{S}_{1t}^{(x)}, S_{1t}^{(y)},
\dot{S}_{1t}^{(y)},\cdots,S_{mt}^{(x)},\dot{S}_{mt}^{(x)},
S_{mt}^{(y)}, \dot{S}_{mt}^{(y)}  )^T$,
$v(t)=(v_{1t}^T,\cdots,v_{mt}^T)^T$ and   $v_{it}$ denotes  the
state disturbance for the $i$th target with  $v_{it} \sim N_4(0,Q)$
and $ Cov(v_{it},v_{jt})=\Sigma_{4\times 4} $ for $i\neq j$.
Consequently $Var(v(t))=(\mathbf{1}_m\otimes\mathbf{1}_m)\Sigma+\{_d
\ Q-\Sigma\}_{i=1}^m$ where $\mathbf{1}_m$ denotes the
$m$-dimensional vector with entry one.
 Furthermore for $\theta_i\  (i=1,\cdots,m)$, we assume
\begin{eqnarray}\label{mixed equation}
\theta_i=\mu_{\theta}+b_i\sim {\rm i.i.d.}N(\mu_{\theta},D),
\end{eqnarray}
where  $\mu_{\theta}^T\triangleq(\alpha,\beta,\gamma)$ represents
the fixed effects and  $b_i\sim N_3(0,D)$ the random effects. In
matrix form we have $ \theta=\mathbf{1}_m\otimes\mu_{\theta}+b$
where $b^T=(b_1^T,\cdots,b_m^T)\sim N_{3m}\left(0,\{_d\
D\}_{i=1}^m\right)$. Model (\ref{state equation})$\sim$(\ref{mixed
equation}) constitute the state equations for MESSM. The measurement
model is more complex and we refer to Khan et al (2005), Pang et al
(2008, 2011) for more details in this respect. These state equations
are meaningful generalization of the present group target tracking
models.
\subsection{Time-dependent  effects}\label{time dependent effects}
In the previous sections both the fixed effects  $a$ and random
effects $b_i$ are assumed to be static, i.e.,  constant across the
time range. In some situations, as can be seen in Example 2 in the
next section, $a$ and $b_i$ can be time-dependent. It turns out that
the results given in previous sections can be easily adapted to
accommodate the time-dependent effects. For the illustrative
purpose, consider the case in which there exists a time point
$1<T'<T$ that for $1\leq t\leq T'$ we have $\theta_i=\Psi_i^{(1)}
a_1+b_{i1}$ with $b_{i1}\sim N(0,D_1)$; while for $T'<t\leq T$ we
have $\theta_i=\Psi_i^{(2)} a_2+b_{i2}$ with $b_{i2}\sim N(0,D_2)$.
For ease of exposition, we assume the individuals are independent
with each other. The unknown parameters include
$\Delta=(a_1,a_2,\delta^T)^T$ where $\delta$ denotes the unknown
parameter contained in $D_1, D_2, Q$ and $R$. In this situation, the
intermediate quantity of  EM algorithm can be shown to be
\begin{eqnarray*}\label{extended Q}
Q(\Delta,\Delta^{\star}) &=& -\frac{m}{2}\log
|D_1(\delta)|-\frac{m}{2}\log |D_2(\delta)| -\frac{Tm}{2}\log
|R(\delta)|-\frac{Tm}{2}\log |Q(\delta)|\nonumber\\
&&-\frac{1}{2}\sum_{i=1}^m {\rm
tr}\left[D_1(\delta)^{-1}\left\{(\Psi_i^{(1)}(a^{\star}_1-a_1)+b_{i1|T})\right.\right.\\
&&\quad\quad \quad\quad
\left.\left.\times(\Psi_i^{(1)}(a^{\star}_1-a_1)+b_{i1|T})^T+{\rm
Var}(b_{i1}|y_{i,1:T},\Delta^{\star}) \right\}\right]\nonumber\\
&&-\frac{1}{2}\sum_{i=1}^m {\rm
tr}\left[D_2(\delta)^{-1}\left\{(\Psi_i^{(2)}(a^{\star}_2-a_2)+b_{i2|T})\right.\right.\\
&&\quad\quad \quad\quad
\left.\left.\times(\Psi_i^{(2)}(a^{\star}_2-a_2)+b_{i2|T})^T +{\rm
Var}(b_{i2}|y_{i,1:T},\Delta^{\star}) \right\}\right]\nonumber
\end{eqnarray*}
\begin{eqnarray*}
 &&-\frac{1}{2}\sum_{i=1}^m \sum_{t=1}^T
{\rm tr}\left[R(\delta)^{-1}\left\{w_{t|T}w_{it|T}^T+{\rm
Var}(w_{it}|y_{1:T},\Delta^{\star})\right\}\right]\nonumber\\
&&-\frac{1}{2}\sum_{i=1}^m \sum_{t=1}^T {\rm
tr}\left[Q(\delta)^{-1}\left\{v_{it|T}v_{it|T}^T+{\rm
Var}(v_{it}|y_{1:T},\Delta^{\star})\right\}\right],\nonumber
\end{eqnarray*}
where $b_{i1|T}, b_{i2|T},w_{i|T}, v_{i|T}$ have the same
explanation as $b_{i|T}, w_{i|T}, v_{i|T}$  in  section \ref{model
estimation}. As for  quasi-Newton algorithm, the score vector now
can be shown to be
\begin{eqnarray*}
\left.\frac{\partial \log L(\Delta | y_{1:T})} {\partial
a_1}\right|_{\Delta=\Delta^{\star}}=\sum_{i=1}^m
\psi_i^TD_1(\delta^{\star})^{-1}b_{i1|T},  \end{eqnarray*}
\begin{eqnarray*} \left.\frac{\partial \log L(\Delta | y_{1:T})}
{\partial a_2}\right|_{\Delta=\Delta^{\star}}=\sum_{i=1}^m
\psi_i^TD_2(\delta^{\star})^{-1}b_{i2|T},
\end{eqnarray*}
\begin{eqnarray*}
 &&\left.\frac{\partial \log L(\Delta |
y_{1:T})}{\partial \delta_j}\right |_{\Delta=\Delta^{\star}}=
-\frac{1}{2}\sum_{i=1}^m {\rm tr } \left[
D_1(\delta^{\star})^{-1}\frac{\partial D_1(\delta^{\star})}{\partial
\delta_j}\right.\\
&&\quad\quad\quad\quad
-D_1(\delta^{\star})^{-1}\left\{b_{i1|T}b_{i1|T}^T+Var(b_{i1}|y_{1:T},
\Delta^{\star})\right\} \left.D_1(\delta^{\star})^{-1}\frac{\partial
D_1(\delta^{\star})}{\partial \delta_j}\right]\nonumber\\
&&-\frac{1}{2}\sum_{i=1}^m {\rm tr } \left[
D_2(\delta^{\star})^{-1}\frac{\partial D_2(\delta^{\star})}{\partial
\delta_j}\right.\\
&&\quad\quad\quad\quad
-D_2(\delta^{\star})^{-1}\left\{b_{i2|T}b_{i2|T}^T+Var(b_{i2}|y_{1:T},
\Delta^{\star})\right\}\nonumber \times
\left.D_2(\delta^{\star})^{-1}\frac{\partial
D_2(\delta^{\star})}{\partial \delta_j}\right]\\
&& +\frac{1}{2}\sum_{i=1}^m \sum_{t=1}^T {\rm
tr}\left[\tilde{E}\left\{e_t(\theta)
e_t(\theta)^T-D_t(\theta)\right\}\frac{\partial
R(\delta^{\star})}{\partial \delta_j}\right]\nonumber\\
&&+\frac{1}{2}\sum_{i=1}^m \sum_{t=1}^T {\rm
tr}\left[\tilde{E}\left\{r_{t-1}
(\theta)r_{t-1}(\theta)^T-N_{t-1}(\theta)\right\}\frac{\partial
Q(\delta^{\star})}{\partial \delta_j}\right]. \nonumber
\end{eqnarray*}
\par Liu et al (2011) used time-dependent effects to model the
dynamics of load of HIV in vivo. Their  model can be formulated as
that defined in (\ref{autoregression1})$\sim$(\ref{autoregression2})
with the modification that  for $1\leq t\leq T'$,
$\theta_i=\mu_{\theta_1}+b_{i1}$ with $b_{i1}\sim N(0,\delta_3)$;
while for $T'<t\leq T$, $\theta_i=\mu_{\theta_2}+b_{i2}$ with
$b_{i2}\sim N(0,\delta_4)$. For this model,  recursive formulas for
EM and quasi-Newton algorithm can be derived out straightforwardly
from those expressions given above. It turns out these formulas  are
similar to those given in section \ref{model estimation} and so the
details are omitted.
\section{Numerical  Studies}\label{numerical studies}
In this section we investigate the performance of  the proposed
algorithms by two numerical examples. The first  example uses the
simulated data which  is generated from the  autoregressive
 with noise model; The second example involves  a  clinical trial
data set which had been   investigated by several  other authors.
For parameter estimation both the EM and BFGS algorithms will be
carried out while only the results of BFGS will be reported because
of the similarity of the results. The variances are calculated from
the observed information matrix based on MKF-KS and the algorithm 3
in Nemeth et al (2013).
\par {\bf Example 1.}\   Consider the  model
 given by
(\ref{autoregression1})$\sim$ (\ref{autoregression2}).  The unknown
parameters include
$\Delta=(\mu_{\theta},\delta_1,\delta_2,\delta_3)$. To generate the
simulated data,  the true parameters  are set to be
$\Delta_0=(0.3,0.3,3,0.1)$; initial state satisfies $x_0\sim N(0,
3.2)$. We only consider the problem of   parameter estimation in
this example and  three sample sizes, $m=15,30,50$ will be
investigated. In  each case, three kinds of  time series,
$T=10,20,30$, are considered. The repetition for each combination is
set to be 500. The number of the random samples generated from the
posterior distribution of random effects is set to be $M=200$.   The
results  are reported in Table 1 which include the parameter
estimates and the corresponding standard errors. From Table 1 it can
be seen that the proposed inference approaches can provide the
reasonable estimates for the unknown parameters.
\par {\bf Example 2.}\  A data set from the  clinical trial of AIDS
 had been investigated in Liu et al (2011), Wu and Ding
(1999) and Lederman et al (1998).  This data set contains the
records of 48 HIV infected patients who are treated with potent
antiviral drugs. Dynamic models with mixed effects for this data set
had been constructed in
 literatures, see Wu and Ding (1999), Liu et al (2011).
In particular  the model proposed in Liu et al (2011)  is just the
model given in the last paragraph in section \ref{time dependent
effects}. For parameter estimation,  they investigated the EM
algorithm and Baysian method. For state estimation, they took the
estimates as the true values of the parameters and then employed the
Kalman filter to estimate the state.
 Here the same model will be investigated and the focus is put on the
statistical inference of such model. The observations $y_{it}$'s are
the base 10 logarithm of the viral load for patient
 $i$ at week $t$. Unknown parameters include $\Delta=(\mu_{\theta_1},
\mu_{\theta_2}, \delta_1,\delta_2, \delta_3,\delta_4)^T$.
 Note  for each  patient, there exist
 some time points that the corresponding records   $y_{it}$'s are missing.
  Thus  the models  in  section \ref{incomplete observations}
   and \ref{time dependent effects} need to be combined together to
   analyze this data set.
   \par For parameter estimation   the results  are reported in Table 2 which include the parameter
estimates and the corresponding standard errors. Table 3 presents
the  estimated individual parameters using the particles
$\{(\theta^{(j)}_t,w_t^{(j)}), j=1,\cdots,M\}$  generated by MKF-KS
algorithm at the last time point. These estimates are just the
weighted means of $\theta_t^{(j)}$ with weights $w_t^{(j)}$. With
the estimated population parameters in hand, the state estimation is
carried out using the MKF-KS algorithm. The resulted filter estimate
and the one-step ahead prediction are plotted in Figure 1 for four
patients who have the most observations among these 48 patients. For
the purpose of comparison  we also run Kalman filter with the
individual parameters replaced by their estimates.  Figure 2
presents the box plots of  mean squared errors of MKF-KS and Kalman
filter for 48 patients. It seems that these two MSE's are similar in
magnitude. This can be explained as follows. On the one hand, Kalman
filter uses all the observations and should outperform MKF-KS
algorithm which only uses the observations up to the present time
point. On the other hand the predicted random effects are taken as
the true random effects in Kalman filter which will results in bias
in state estimation. While for MKF-KS the random effects are
integrated out when the states are estimated and so less affected by
estimating errors. Both factors affect   the magnitude of the MSE's.
Recall contrary to Kalman filter  the main advantages of MKF-KS is
that without the known random effects it also can provide the
recursive state estimation. This point is more important in the
setting of sparse data in which the random effects can not be
estimated accurately.
\section{Conclusion}\label{conclusions}
We consider both the parameter  and state estimation for the linear
mixed-effects state space model which can accommodate the correlated
individuals. For parameter estimation   EM and score based
algorithms are investigated based on disturbance smoothing. The
implementation of
 EM and score based algorithms only require the random samples of random
 effects from the posterior distribution. Particularly the proposed  EM algorithm can be regarded as  a
Rao-Blackwellized  version of that proposed in Liu et al (2011). For
state estimation, because longitudinal data set usually involves
sparse data with which  random effects can not be estimated
accurately,  we advocate state estimation should be carried out
without assuming the random effects being known.
 To this end a  kernel smoothing based mixture Kalman filter is
 proposed to estimate the state. Numerical studies show the proposed
 inferences  perform  well in the setting of  finite
 samples.
 The proposed models and statistical
 inferences can be extended by different ways. For example
nonlinear  mixed-effects  state space model with additive  Gaussian
error can be
 handled by the similar ideas in this paper without much difficulty. But for the
 general nonlinear/non-Gaussian state space model with
 mixed-effects, the proposed algorithms can not apply and
 new inference techniques need to be developed. Another interesting problem
 is how to carry out the parameter estimation in a recursive manner. For the ordinary
  fixed-effect state space models, there have existed  some studies  in
  this respect. Extending such inferences to state
  space model with mixed effects also is meaningful.

\newpage
\appendix
\appendix
\begin{table}
\caption{Parameter estimates and standard errors with true parameter
$\mu=0.3,\delta_1=0.3, \delta_2=3,\delta_3=0.1$.} \label{table1}
\centering
\begin{tabular}{rccccccccc}
\hline\hline
&&\multicolumn{2}{c}{$\mu_{\theta}$}&\multicolumn{2}{c}{$\delta_1$}
&\multicolumn{2}{c}{$\delta_2$}&\multicolumn{2}{c}{$\delta_3$}\\
 Cases&&Estimate&SE&Estimate&SE&Estimate&SE&Estimate&SE\\
 \hline
$m$=\small{15}&$T$=\small{10}&\small{0.27}&\small{0.05}&\small{0.40}&\small{0.1}
&\small{4.70}&\small{0.52}&\small{0.14}&\small{0.007}\\
&$T$=\small{20}&\small{0.26}&\small{0.04}&\small{0.38}&\small{0.08}&\small{3.21}
&\small{0.47}&\small{0.07}&\small{0.006}\\
&$T$=\small{30}&\small{0.28}&\small{0.04}&\small{0.34}&\small{0.03}
&\small{3.17}&\small{0.27}&\small{0.07}&\small{0.006}\\
\hline
$m$=\small{30}&$T$=\small{10}&\small{0.27}&\small{0.02}&\small{0.37}&\small{0.07}
&\small{3.82}&\small{0.29}&\small{0.13}&\small{0.005}\\
&$T$=\small{20}&\small{0.28}&\small{0.02}&\small{0.34}&\small{0.03}&\small{2.43}&\small{0.17}&\small{0.12}&\small{0.006}\\
&$T$=\small{30}&\small{0.31}&\small{0.01}&\small{0.24}&\small{0.02}&
\small{2.71}&\small{0.20}&\small{0.08}&\small{0.005}\\
\hline $m$=\small{50}&$T$=\small{10}&\small{0.30}&\small{0.01}&\small{0.34}&\small{0.06}&\small{3.51}&\small{0.27}&\small{0.12}&\small{0.004}\\
&$T$=\small{20}&\small{0.31}&\small{0.01}&\small{0.32}&\small{0.04}&\small{3.22}
&\small{0.21}&\small{0.11}&\small{0.005}\\
&$T$=\small{30}&\small{0.31}&\small{0.01}&\small{0.32}&\small{0.04}&\small{2.87}&\small{0.20}&\small{0.11}&\small{0.001}\\
\hline
\end{tabular}
\end{table}

\begin{table}
\caption{Population parameter estimates and standard errors}
\label{table2} \centering
\begin{tabular*}{0.8\textwidth}{@{\extracolsep{\fill}}lcccccc}
\hline \hline
&$\mu_{\theta_1}$&$\mu_{\theta_2}$&$\delta_1$&$\delta_2$&$\delta_3$&$\delta_4$\\
\hline
Estimates&\small{0.85}&\small{0.86}&\small{0.33}&\small{0.76}&\small{0.007}&\small{0.044}\\
\hline
SE's&\small{0.06}&\small{0.04}&\small{0.08}&\small{0.23}&\small{0.002}&\small{0.01}\\
\hline
\end{tabular*}
\end{table}
\clearpage
\begin{table}
\caption{Estimation of the individual parameters for 48 patients}
\begin{tabular}{lccccccc}
\hline\hline
$\theta_{i}^{(1)}:$&&&&&&&\\
\small{0.868915}&\small{0.849607}&\small{0.868957}
&\small{0.827189}&\small{0.851339}&\small{0.847058}&\small{0.848824}&\small{0.851733}\\
\small{0.849319}&\small{0.868136}&\small{0.838490}&\small{0.835906}&\small{0.859012}&\small{0.825839}&\small{0.846816}&\small{0.859276}\\
\small{0.867417}&\small{0.857401}&\small{0.843068}&\small{0.835888}&\small{0.837270}
&\small{0.852747}&\small{0.832048}&\small{0.842219}\\
\small{0.850987}&\small{0.852832}&\small{0.835151}
&\small{0.856031}&\small{0.872748}&\small{0.873013}&\small{0.840243}&\small{0.851437}\\
\small{0.893272}&\small{0.865324}&\small{0.853658}&\small{0.858038}
&\small{0.863467}&\small{0.836726}&\small{0.837801}&\small{0.846284}\\
\small{0.809998}&\small{0.844643}&\small{0.846764}&\small{0.848282}
&\small{0.846723}&\small{0.833354}&\small{0.837123}&\small{0.828165}\\
\hline
$\theta_{i}^{(2)}:$&&&&&&&\\
\small{0.852200}&\small{0.841703}&\small{0.947872}&\small{0.894520}&\small{0.848136}
&\small{0.975277}&\small{0.862129}&\small{0.925632}\\
\small{0.859690}&\small{0.865170}&\small{0.762160}&\small{0.921414}
&\small{0.932375}&\small{0.892159}&\small{0.858365}&\small{0.776656}\\
\small{0.938613}&\small{0.870105}&\small{0.900053}&\small{0.844046}
&\small{0.962848}&\small{0.872085}&\small{0.826155}&\small{0.900559}\\
\small{0.843995}&\small{0.712202}&\small{0.901383}&\small{0.924811}&\small{0.910035}
&\small{0.957107}&\small{0.920373}&\small{0.878261}\\
\small{0.868421}&\small{0.928374}&\small{0.867860}&\small{0.915143}&\small{0.849401}
&\small{0.908050}&\small{0.944003}&\small{0.925122}\\
\small{0.936315}&\small{0.905399}&\small{0.872215}&\small{0.858642}&\small{0.821628}
&\small{0.875818}&\small{0.753861}&\small{0.948502}\\
\hline
\end{tabular}
\end{table}
\clearpage
\begin{figure}[h]
\centering
\includegraphics[width=1.1\textwidth,height=2.5in]{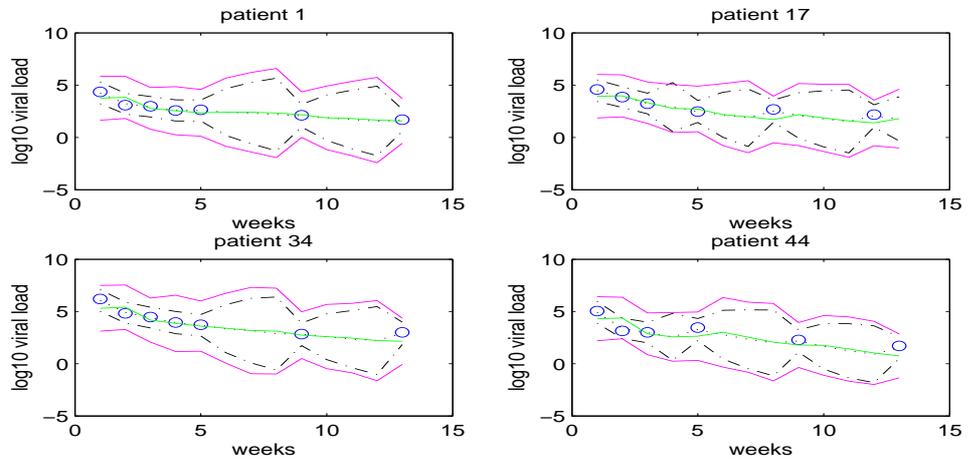}
\caption{\footnotesize{Estimation  of viral load for four patients
in the HIV dynamic study. The circles represent base 10 logarithm of
the viral loads.  The green solid lines represent the one-step ahead
prediction; The dotted  lines represent the filtering estimates; The
dashed lines represent the 95\% confidence interval  of the
filtering estimates; The pink solid lines represent the 95\%
confidence interval  of the one-step ahead prediction.}}
\label{fig1}
\end{figure}

\begin{figure}[h]
\centering
\includegraphics[width=0.8\textwidth,height=1.8in]{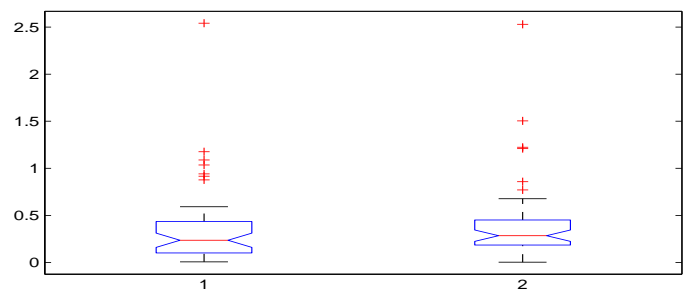}
\caption{\footnotesize{Mean square errors of the one-step ahead
prediction  for 48 patients. The left panel corresponds to the
MKF-KS algorithm. The right panel corresponds to the Kalman filter
with the estimated individual parameters.}} \label{fig2}
\end{figure}
\end{document}